\documentstyle[11pt,newpasp,twoside,epsf]{article}
\def\edcomment#1{\iffalse\marginpar{\raggedright\sl#1\/}\else\relax\fi}
\reversemarginpar

\begin{document}
\title{New Binary and Millisecond Pulsars from Arecibo Drift-Scan Searches}
 \author{M. A. McLaughlin$^{1}$, D. R. Lorimer$^{1}$, D. J. Champion$^{1}$, \\
 Z. Arzoumanian$^{2}$,  D. C. Backer$^{3}$, J. M. Cordes$^{4}$, \\
A. S. Fruchter$^{5}$,  A. N. Lommen$^{6}$\& K. M. Xilouris$^{7}$}
\affil{$^{1}$University of Manchester,
Jodrell Bank Observatory, Macclesfield, Cheshire, SK11~9DL, UK,
$^{2}$Laboratory for High-Energy Astrophysics, USRA/LHEA,
NASA Goddard Space Flight
Center, Code 662, Greenbelt, MD 20771, USA,
$^{3}$Astronomy Department, University of California, Berkeley, CA 94720, USA,
$^{4}$Astronomy Department, Cornell University, Ithaca, NY 14853, USA,
$^{5}$Space Telescope Science Institute, 3700 San Martin Drive, Baltimore,
MD 21218, USA,
$^{6}$Department of Physics \& Astronomy, Franklin \& Marshall College,
PO Box 3003, Lancaster, PA 17604, USA,
$^{7}$University of Virginia, Department of Astronomy, PO Box 3818,
Charlottesville, VA 22903, USA}

\begin{abstract}
We discuss four recycled pulsars found in Arecibo drift-scan
searches. PSR~J1944+0907 has a spin period of 5.2 ms and is
isolated. The 5.8-ms pulsar J1453+19 may have a low-mass companion.
The isolated 56-ms pulsar J0609+2130 is possibly the remnant of a
disrupted double neutron star binary. The 41-ms pulsar J1829+2456 is
in a relativistic orbit. Its companion is most likely another
neutron star.
\end{abstract}

\section{Introduction}

Recycled pulsars are believed to be formed when an old neutron star is
spun up through the accretion of matter from a binary companion
(Bisnovatyi-Kogan \& Komberg 1974). For those companions massive
enough to explode as a supernova, the binary lifetime is relatively
short ($\sim 10^{6-7}$ yrs) and the most likely outcome is the
disruption of the binary. Those systems which survive the explosion
are the double neutron star (DNS) binaries.  For less massive
companions, where the period of spin-up is longer ($\sim 10^8$ yr),
the collapse leaves a white dwarf star in orbit around a rapidly
spinning millisecond pulsar (MSP).  Of the roughly 1700 known radio
pulsars, only 100 are MSPs and less than 10 are in DNS systems.  In
order to understand the population of recycled pulsars in detail, a
larger sample is required.  Here we describe the implications for
four recycled pulsars recently detected in Arecibo drift-scan searches.

\section{Data acquisition and analysis}

The data were taken with the Arecibo telescope at 430 MHz in
drift-scan mode.  A source drifts through the $10\arcmin$ beam in
$\sim$~42~s.  The incoming signals were passed to the Penn State
Pulsar Machine (PSPM) which summed the two independent polarizations
before 4-bit sampling the band into $128\times60$-kHz channels every
80 $\mu$s. The data were written directly to magnetic tape and later
processed using COBRA, Jodrell Bank's 182-node Linux
cluster. Processing consisted of dedispersion over a range of trial
dispersion measures (DMs), Fourier transforms of length $2^{19}$-pts
and a single-pulse search. RFI mitigation was achieved by creating
spectral masks based on consecutive 0-DM time series. For further
details, see Lorimer et al. (2004). Analysis of data
corresponding to roughly 1700 deg$^{2}$ of sky has resulted in the
discovery of 12 new pulsars. We discuss here only the four 
these new pulsars that are recycled.  In Figure~1, we show the pulse
profiles of these pulsars. 

\begin{figure}[h]
\plotone{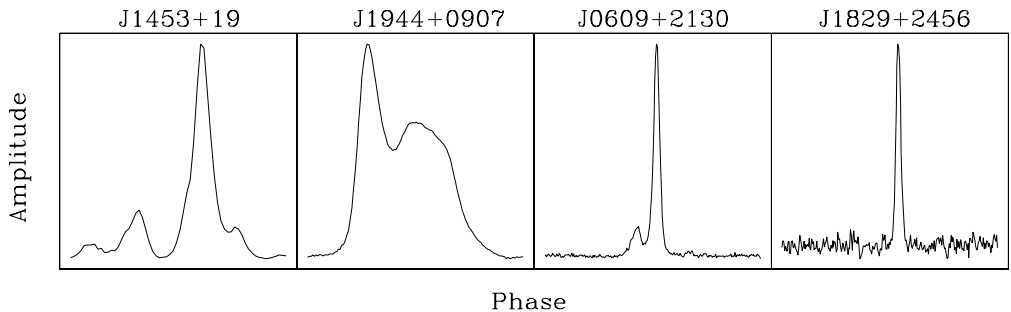}
\caption{Integrated 
pulse profiles for the new pulsars. Each plot shows 360$\deg$
of rotational phase.}
\end{figure}

We are carrying out regular Arecibo timing observations of all the
new pulsars using the PSPM. Table 1 shows a phase-connected timing
solution spanning 500 days
for PSR J1944+0904. Solutions for J0609+2130 and J1829+2456
can be found in Lorimer et al.~(2004) and Champion et al.~(2004)
respectively. We do not currently have a solution for PSR~J1453+19.

\begin{center}
\begin{table}[h!]
\caption{\label{tab:timing}Observed and derived parameters for J1944+0907}

\begin{tabular}{ll}
\hline
\hline
Parameter & Value \\
\hline
Right ascension (h:m:s) (J2000) \hspace{1in} & 19:44:09.321(4) \\
Declination (deg:m:s) (J2000) & 09:07:23.25(1) \\
Spin period, $P$ (ms) &	 5.185201903773(2)\\
Epoch of period (MJD) &	 52845.0 \\
Timing data span (MJD) & 52595--53093\\
Period derivative, $\dot{P}$ ($\times 10^{-20}$ s s$^{-1}$) & 1.65(5) \\
Dispersion measure, DM (cm$^{-3}$ pc) & 24.093(5) \\
\hline
Characteristic age, $\tau$ (Gyr) & 5.0\\
Magnetic field strength, $B$ ($10^8$ G) & 3.0\\
Distance, $d$ (kpc) & 1.8 \\
\hline
\end{tabular}
\hspace{2in}
The numbers in parentheses represent uncertainties
in the least significant digit. The characteristic age
and magnetic field are calculated by the standard Manchester \& 
Taylor (1977) formulae.
\end{table}
\end{center}

\section{Two new millisecond pulsars}

We discovered two MSPs in our searches.  While we do not yet have a
phase connection for J1453+19, we can certainly rule out a ``regular''
0.3 M$_{\odot}$ white-dwarf companion. Further observations of the
pulsar are underway. The isolated 5.8-ms pulsar J1944+0907
scintillates strongly at 430 MHz (see Figure~2), with scintillation
bandwidth and timescale of $\Delta\nu$ = 180~kHz and $\Delta t$ =
101~s. For its DM-derived (i.e. Cordes \& Lazio 2002) distance of
1.8~kpc, this implies a transverse velocity of 114 km~s$^{-1}$.

\begin{figure}[h]
\plotone{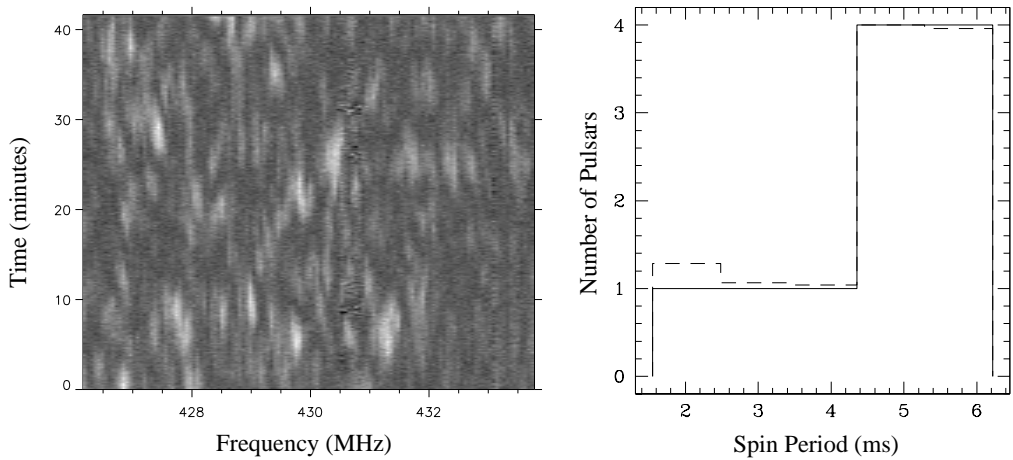}
\caption{Left: Dynamic spectrum (i.e. on-pulse intensity versus time
and frequency) for J1944+0907 at 430~MHz. The spectrum was formed with
10-s integrations and using the PSPM, with 128 frequency channels
across the 7.68-MHz bandwidth.  Right: Solid line shows the
distribution of spin period for MSPs detected in Arecibo drift-scan
searches.  The dashed line shows the distribution corrected for
sensitivity losses at short periods.}
\end{figure}

As shown in Table~2, including J1944+0907, and possibly J1453+19,
there are now 14(15) 
isolated MSPs in the Galactic disk. We do not include isolated MSPs
in globular clusters in this table as their formation processes may be
quite different. How have these isolated MSPs in the disk formed?  It
is possible that they are formed through the accretion induced
collapse of white dwarfs (Bailyn~\& Grindlay 1990) or through white
dwarf mergers (Michel 1987). Another possibility is that the
companions of these pulsars have been destroyed by ablation or tidal
disruption (Bhattacharya \& van den Heuvel 1991, Radhakrishnan \&
Shukre 1986). If this is the case, we might expect isolated MSPs to
have higher velocities than binaries because closer binaries are more
likely to suffer from these effects, and closer binaries are expected
to have average higher velocities (Tauris \& Bailes 1996).
Measurements of isolated MSP velocities are therefore important for
determining their origin.  
With DM-derived distances of 1.2 and 1.8~kpc for J1453+19
and J1944+0907, respectively, proper motion measurements may be
possible with long-term timing. At our current level of precision,
we should be able to confirm our scintillation speed measurement
for J1944+0907 by the end of 2005.

In Table~2, we list transverse velocities
for isolated MSPs inferred from timing proper motion or scintillation
measurements (Toscano 1999, Gothoskar \& Gupta 2000, Nicastro et
al. 2001).   The average velocity of isolated MSPS
is roughly 70 km s$^{-1}$, slightly lower than the average velocity of
the binary millisecond pulsar population (e.g. Nicastro et al. 2001)
and in contrast to what is expected if these isolated MSPs have
ablated or tidally disrupted their companions.  However, because the
velocity determinations for MSPs depend on their largely uncertain
distances, this conclusion is tentative.

\begin{center}
\begin{table}[h]
\caption{Isolated millisecond pulsars.
Distances are derived from Cordes \& Lazio (2002). 
Scintillation-derived velocities are italicized.}
\begin{tabular}{lcccc}
\hline
\hline
Pulsar & P (ms) & DM (pc cm$^{-3}$) & D (kpc) & $V_t$ (km s$^{-1}$) \\
\hline
J0030$+$0451 \hspace{0.5in} & 4.87  & 4.33 & 0.3 & $<$ {\it 15} \\
J0711$-$6830 & 5.49 & 18.4 & 0.9 & 87 \\
J1024$-$0719 & 5.16 & 6.49 & 0.4 & 124 \\
J1453$+$19 & 5.79 & 14.2  & 1.2 & \\
J1629$-$6902 & 6.00 &  29.5 & 1.0 & \\
J1709$+$2313 & 4.63 & 26 & 1.4 & 89\\
J1721$-$2457 & 3.50 &  47.8 & 1.3 & \\
J1730$-$2304 & 8.12& 9.61 & 0.5 & {\it 56} \\
J1744$-$1134 & 4.08 & 3.14 & 0.4 & 36 \\
J1843$-$1113 & 1.85 & 59.96 & 2.0 &    \\
J1905$+$0400 & 3.78 & 25.7  & 1.3  &    \\
B1937$+$21 & 1.56 & 71.0 & 3.6 & 79 \\
J1944$+$0907 & 5.19 & 24.1 & 1.8 & {\it 114} \\
J2124$-$3358 & 4.93 & 4.62 & 0.3 & 53 \\
J2322$+$2057 & 4.81 & 13.4 & 0.8  & 80 \\
\hline
\end{tabular}
\end{table}
\end{center}

In total, Arecibo drift-scan surveys have detected 11 pulsars (new and
previously known) with periods less than 10~ms (Ray et al. 1995,
Camilo et al. 1996, Ray et al. 1996, Lommen et al. 2000).  In
Figure~2, we show the distribution of spin periods along with a
distribution corrected for the loss of sensitivity at short spin
periods due to dispersion, scattering, and our finite sampling
interval. If we naively assume that the actual distribution of spin
periods is uniform and apply the analysis of Chakrabarty et
al. (2003), we find that the minimum spin period of this distribution
(at 95\% confidence) is 1.2~ms. In their analysis of millisecond
oscillations from X-ray bursters, Chakrabarty et al. found a minimum
spin period of 1.3~ms. These limiting periods are much higher than
those predicted by standard neutron star equations of state (e.g. Cook
et al. 1994), suggesting that some other mechanism limits the minimum
spin period.  One possibility is that gravitational radiation can
carry away angular momentum from accreting neutron stars (Wagoner
1984, Bildsten 1998).

\section{New recycled pulsars}

In Lorimer et al.~(2004), we proposed that J0609+2130 is a disrupted
double neutron star binary.  Simulations
(e.g. Portegies Zwart \& Yungelson 1998) show that, given standard
evolutionary scenarios and kick velocities, we expect roughly 10
disrupted systems to every DNS.  Therefore, why do we not observe more
pulsars like J0609+2130?  One reason may be that the standard
assumptions used in simulations about evolutionary scenarios and kick
velocities are incorrect.  Measuring proper motions of recycled
pulsars may provide insights into their evolutionary histories.
We expect to measure or constrain the proper motion of 
PSR~J0609+2130 through our on-going timing observations.

\begin{figure}[h]
\plotone{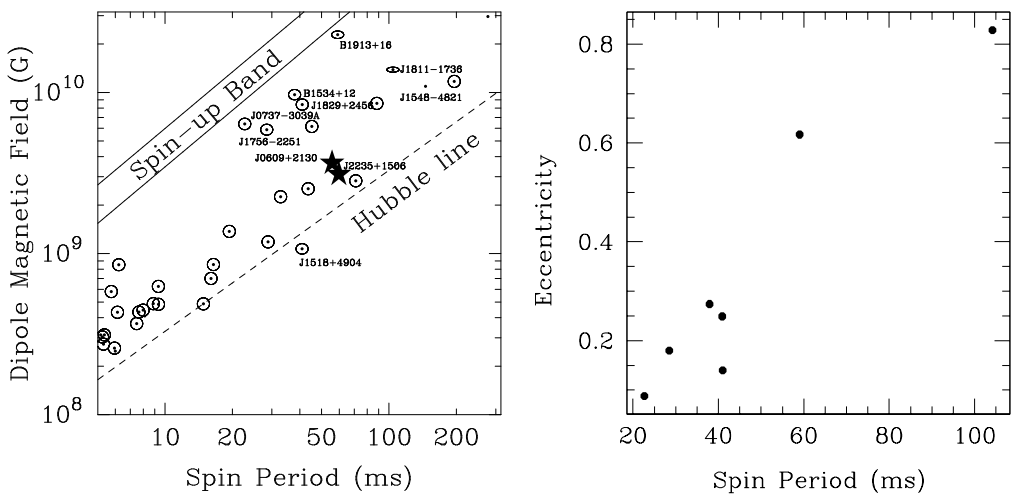}
\caption{Left: Part of the $B-P$
diagram highlighting the unique positions (starred symbols) of the
solitary pulsars J0609+2130 and J2235+1506 (Camilo, Nice \& Taylor
1993).  Pulsars with surrounding ellipses are members of binary
systems. Known DNS binary systems, including likely new DNS systems
J1756--2251 and J1829+2456, are labeled.  The ``Hubble line'' is the
locus of points of constant characteristic age equal to a Hubble time
(assumed to be 15 Gyr). The spin-up band is the end point of spin-up
by accretion for the model of Arzoumanian, Cordes \& Wasserman (1999).
Right: Eccentricity plotted against spin period for the double neutron
star binaries (again including J1756-2251 and J1829+2456.) Perhaps this is because 
high
eccentricity systems with compact orbits (and hence small post-accretion spin periods)
will quickly inspiral and become undetectable while high eccentricity systems with wider orbits 
(and larger post-accretion spin periods) will inspiral more slowly and remain detectable for longer.}
\end{figure}

The 41-ms pulsar J1829+2456 is in a relativistic 28-hr orbit of
eccentricity 0.13 (Champion et al. 2004).  From the measurement
of periastron advance, we find the total mass of the system to
be 2.4 M$_{\odot}$ and constrain the minimum companion mass to be 1.2
M$_{\odot}$.  This constraint, along with the similarity of the spin
parameters to those of the double neutron star binaries, makes the
companion likely to be another neutron star. We have searched our data
for a companion radio pulsar by correcting the time series for the
known pulsar orbit but thus far have detected no companion signals. A
sensitive optical search to rule out a massive white dwarf companion
is planned. If J1829+2456 is indeed a DNS, it will be the eighth such
system known.  With the large number of systems now known, it is
finally possible to look for trends in system properties which may
shed light on the evolution of these objects. For instance, we show in
Figure~3 a possible correlation between eccentricity and spin period.

\acknowledgements

Many thanks to the organizers for putting together such a stimulating
and entertaining conference.  The Arecibo observatory, a facility of
the National Astronomy and Ionosphere Center, is operated by Cornell
University in a co-operative agreement with the National Science
Foundation (NSF).  We thank Alex Wolszczan for making the PSPM freely
available for use at Arecibo, Paulo Friere for observing assistance
and Ingrid Stairs, Joe Taylor and Joel Weisberg for graciously
donating valuable snippets of observing time.


\begin{references}
\reference{Arzoumanian,~Z., Cordes,~J.~M., Wasserman,~I. 1999, ApJ, 520, 696}
\reference{Bailes, M., et al.\ 
1997, \apj, 481, 386} 
\reference{Bailyn, C.~D.~\& 
Grindlay, J.~E.\ 1990, \apj, 353, 159} 
\reference{Bhattacharya, D.~\& van den Heuvel, E.~P.~J.\ 1991, Phys. Rev., 203, 1} 
\reference{Bildsten, L.\ 1998, \apjl, 
501, L89 }
\reference{Bisnovatyi-Kogan,~G.~S. \& Komberg,~B.~V. 1974, Sov. Astron., 18, 217}
\reference{Camilo, F., Nice, D.~J. \& Taylor, J.~H. 1993, ApJ, 412, L37}
\reference{Camilo, F. et al. 1996, \apj, 
469, 819}
\reference{Chakrabarty, D. et al. 2003, Nature, 424, 42}
\reference{Champion, D. et al. 2004, MNRAS, in press, (astro-ph/0403553)}
\reference{Cook, 
G.~B., Shapiro, S.~L., \& Teukolsky, S.~A.\ 1994, \apj, 424, 823} 
\reference{Cordes,~J.~M. \& Lazio,~T. J.~W. 2003, (astro-ph/0207156)}
\reference{Gothoskar, P.~\& 
Gupta, Y.\ 2000, \apj, 531, 345 }
\reference{Lommen, A.~N. et al. 2000, \apj, 545, 1007}
\reference{Lorimer, D.~R., et al.\ 
2004, \mnras, 347, L21}
\reference{Manchester, 
R.~N.~\& Taylor, J.~H.\ 1977, San Francisco : W.~H.~Freeman}
\reference{Michel, F.~C.\ 1987, Nature, 329, 
310 }
\reference{Nicastro, L. et al. 2001, 2001, \aap, 368, 1055} 
\reference{Portegies~Zwart,~S.~F. \& Yungelson,~L.~R. 1998, A\&A, 332, 173}
\reference{Radhakrishnan, 
V.~\& Shukre, C.~S.\ 1986, \apss, 118, 329} 
\reference{Ray, P.~S. et al.\ 1995, 
\apj, 443, 265}
\reference{Ray, P.~S. et al. 1996, \apj, 470, 1103}
\reference{Tauris, T.~M.~\& 
Bailes, M.\ 1996, \aap, 315, 432}
\reference{Toscano, M. et al. 1999, \mnras, 307, 925}
\reference{Wagoner, R.~V.\ 1984, \apj, 
278, 345 }
\end{references}
\end{document}